\documentclass[twocolumn,showpacs,prd]{revtex4}
\usepackage{mathrsfs,bm,multirow}
\usepackage{longtable,lscape}
\usepackage{txfonts}
\usepackage{amssymb}
\usepackage{indentfirst}
\usepackage{graphicx,,booktabs}
\usepackage{color}
\usepackage{amssymb}

\begin{document}

\title{Simulating the charged charmoniumlike structure $Z_c(4025)$}
\author{Xiao Wang$^{1,2}$}
\author{Yuan Sun$^{1,2}$}
\author{Dian-Yong Chen$^{1,3}$}
\email{chendy@impcas.ac.cn}
\author{Xiang Liu$^{1,2}$\footnote{Corresponding author}}\email{xiangliu@lzu.edu.cn}
\author{Takayuki Matsuki$^4$}
\email{matsuki@tokyo-kasei.ac.jp}
\affiliation{$^1$Research Center for Hadron and CSR Physics,
Lanzhou University $\&$ Institute of Modern Physics of CAS,
Lanzhou 730000, China\\
$^2$School of Physical Science and Technology, Lanzhou University,
Lanzhou 730000, China\\
$^3$Nuclear Theory Group, Institute of Modern Physics, Chinese
Academy of Sciences, Lanzhou 730000, China\\
$^4$Tokyo Kasei University, 1-18-1 Kaga, Itabashi, Tokyo 173-8602,
Japan}

\begin{abstract}

Inspired by recent observation of charged charmoniumlike structure $Z_c(4025)$, we explore the $Y(4260)\to
(D^*\bar{D}^*)^- \pi^+$ decay through the initial-single-pion-emission mechanism, where the
$D^*\bar{D}^*\to D^*\bar{D}^*$ interaction is studied by the ladder approximation including a
non-interacting case. Our calculation of the differential decay width for $Y(4260)\to (D^*\bar{D}^*)^-
\pi^+$
indicates that a charged enhancement structure around $D^*\bar{D}^*$ appears in the $D^*\bar{D}^*$
invariant mass spectrum for this process, which can correspond to newly observed $Z_c(4025)$ structure.

\end{abstract}
 \maketitle

\section{introduction}

Very recently, the BESIII Collaboration announced the observation of a charged charmoniumlike structure
$Z_c(4025)$, which appears in the $\pi^\mp$ recoil mass spectrum of $e^+e^- \to
(D^{*}\bar{D}^{*})^{\pm}\pi^\mp$ at $\sqrt{s}=4.26$ GeV. Its mass and width are $M=(4026.3\pm2.6\pm3.7)$
MeV and $\Gamma=(24.8\pm5.6\pm7.7)$ MeV \cite{Ablikim:2013nra}. Thus, $Z_c(4025)$ is near the
$(D^*\bar{D}^*)^\pm$ threshold.

Before the observation of $Z_c(4025)$, the Belle Collaboration reported a charged bottomoniumlike
structure $Z_b(10650)$ by studying $\Upsilon(10860)\to (B^*\bar{B}^*)^\pm \pi^\mp$ \cite{Adachi:2012cx},
where $Z_b(10650)$ is near the $(B^*\bar{B}^*)^\pm$ threshold. The similarity between $Z_c(4025)$ and
$Z_b(10650)$ indicates that $Z_c(4025)$ can be as the counterpart of $Z_b(10650)$. In Ref.
\cite{Chen:2012yr}, we have proposed an explanation via the initial-single-pion-emission (ISPE) mechanism
why there exists $Z_b(10650)$ in the decay $\Upsilon(10860)\to (B^*\bar{B}^*)^\pm \pi^\mp$.  What is more
important is that we have already predicted a charged structure near the $D^*\bar{D}^*$ threshold in the
$(D^*\bar{D}^*)^\pm$ invariant mass spectrum of $\psi(4415)\to (D^*\bar{D}^*)^\pm \pi^\mp$
in the same paper.

This recent experimental discovery of $Z_c(4025)$ provides us more chances to further reveal the
underlying mechanism behind this novel phenomenon. In the past decades, experimental search for exotic
states beyond the conventional hadron configurations is an important and intriguing research topic.
The peculiarities of $Z_c(4025)$ immediately drive us to recognize that $Z_c(4025)$ can be the most
reliable candidate as an exotic state. However, before giving a one-sided view, we need to exhaust all the
possibilities under conventional frameworks.

Along this way, in this paper we analyze the decay process $Y(4260)\to (D^*\bar{D}^*)^\pm \pi^\mp$ via the
ISPE mechanism or its extension to include higher orders. This mechanism \cite{Chen:2011pv} was first
proposed to understand why two bottomoniumlike structures $Z_b(10610)$ and $Z_b(10650)$ can be found in
the $\Upsilon(nS)\pi^\pm$ ($n=1,2,3$) and $h_b(mP)\pi^\pm$ ($m=1,2$) invariant mass spectra of $e^+e^-\to
\Upsilon(nS)\pi^+\pi^-,\,h_b(mP)\pi^+\pi^-$ at $\sqrt{s}=10865$ MeV \cite{Collaboration:2011gja}. Later,
the ISPE mechanism has been extensively applied to study the hidden-charm dipion/dikaon decays of higher
charmonia and charmoniumlike states \cite{Chen:2011xk,Chen:2013wca,Chen:2013bha}, the hidden-bottom dipion
decays of $\Upsilon(11020)$ \cite{Chen:2011pu}, and the hidden-strange dipion decays of $Y(2175)$
\cite{Chen:2011cj}, where many novel phenomena of charged enhancement structures have been predicted. In
this work, by studying the process $Y(4260)\to (D^*\bar{D}^*)^- \pi^+$, we expect to answer whether a
newly observed charged structure $Z_c(4025)$ can be explained by the ISPE mechanism, which is an
intriguing research topic, too, to search for the underlying mechanism behind this kind of novel
phenomena.

This work is organized as follows. After introduction, we present the calculation of $Y(4260)\to
(D^*\bar{D}^*)^\pm \pi^\mp$ via extension of the ISPE mechanism, where description of the interaction
$D^{*0}{D}^{*-}\to D^{*0}{D}^{*-}$ is given by the ladder diagrams applying an effective Lagrangian
approach. In Sec. \ref{sec3}, the numerical results are shown in comparison with the experimental data.
The Last section is a short summary.

\section{$Y(4260)\to (D^*\bar{D}^*)^- \pi^+$ decay}\label{sec2}

\begin{figure}[htp]
\begin{tabular}{c}
\includegraphics[width=0.28\textwidth]{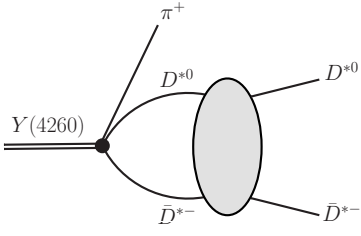}
\end{tabular}
\caption{(color online). The typical diagram depicting $Y(4260)\to
\pi^+ D^{*0}{D}^{*-}$ via the ISPE mechanism. Here, the grey
kernel represents the $D^{*0}{D}^{*-}\to D^{*0}{D}^{*-}$
interaction given in Fig. \ref{sp}.}\label{Fig:main}
\end{figure}

The ISPE mechanism for $Y(4260)\to (D^*\bar{D}^*)^- \pi^+$ is shown by the diagram in Fig.
\ref{Fig:main}. Due to this mechanism, the emitted pion from the $Y(4260)$ decay plays a very important
role and has
continuous energy distribution, which easily enables $D^{*0}$ and $D^{*-}$
with low momenta to interact with each other. Thus, in the
following our main task is to describe the $D^{*0}D^{*-}\to
D^{*0}{D}^{*-}$ interaction and combine this reaction with the
corresponding $Y(4260)$ decay. In this work, we include a tree diagram, i.e.,
the case that the kernel does not include an interaction.

\begin{figure}[htp]
\begin{tabular}{c}
\includegraphics[width=0.28\textwidth]{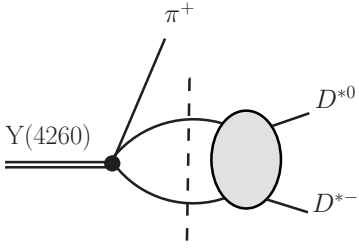}
\end{tabular}
\caption{(color online). The diagram in which a cut is inserted in Fig. \ref{Fig:main}
which is used to calculate the final total amplitude. } \label{Fig:cut}
\end{figure}

\begin{figure}[htbp]
\begin{tabular}{cc}
\scalebox{0.35}{\includegraphics{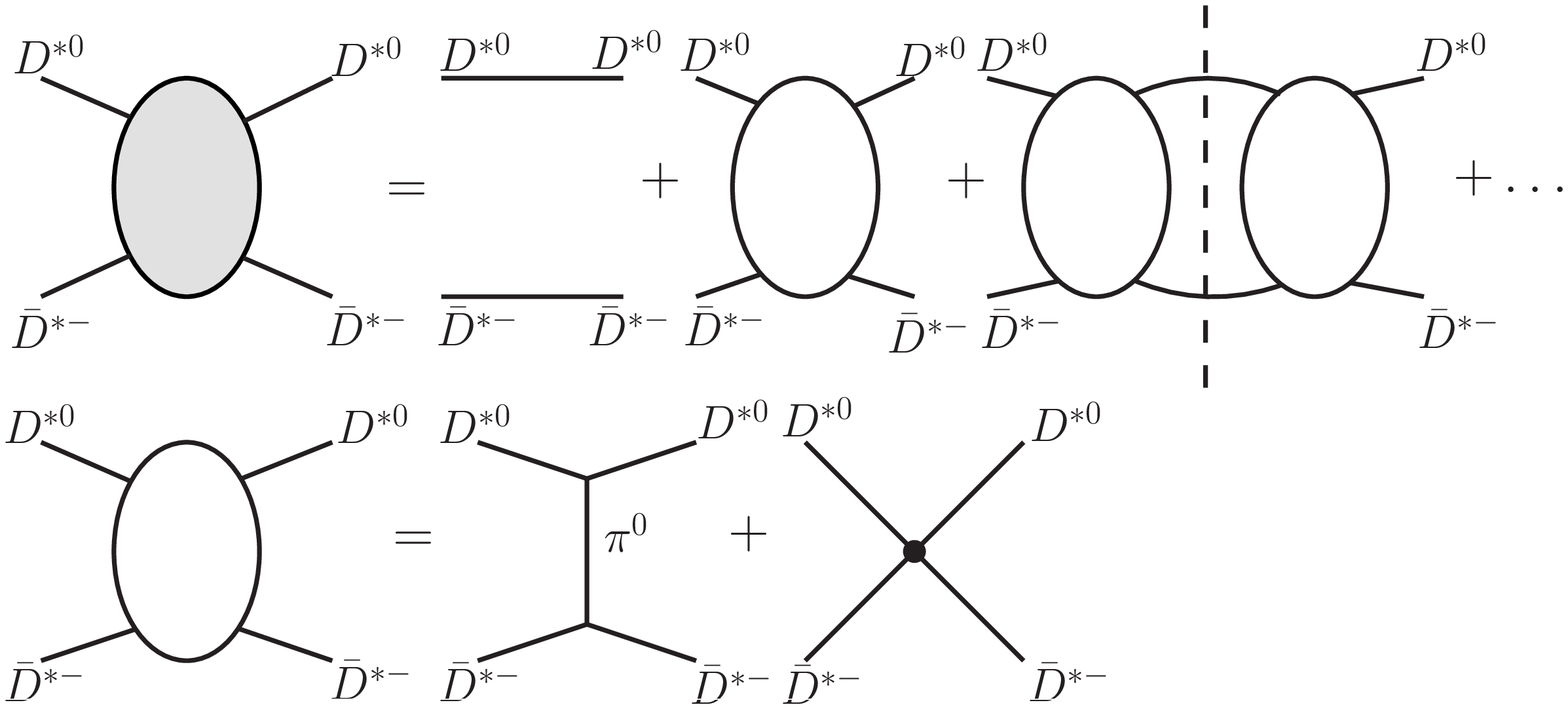}}
\end{tabular}
\caption{The ladder approximation denoted by a grey blob for the $D^{*0}{D}^{*-}\to
D^{*0}{D}^{*-}$ interaction. Here, we consider an elementary contribution of the four-point vertex denoted
by a white blob coming from the pion exchange and the contact term. The vertical dashed line is a cut.}
\label{sp}
\end{figure}

To calculate the $D^{*0}{D}^{*-}\to D^{*0}{D}^{*-}$ interaction
near the threshold as depicted by a grey kernel shown in Fig.
\ref{Fig:main}, we adopt a ladder approximation presented in
Fig. \ref{sp}, {where we borrow some ideas from the Bethe-Salpeter equation \cite{bs}.} After expanding the amplitude by partial wave
bases, the ladder diagrams with $n$ loops can be expressed as a
geometric series, which allows us to sum over all the ladder
diagrams (see the first row in Fig. \ref{sp}). 
{This treatment is allowed when the higher loop contribution is as large as the lower-order one.}
In order to have a geometric series we make a further approximation for the ladder diagrams to insert cuts
in between all the white kernels as shown in the first row of Fig. \ref{Fig:main}.
Furthermore, in
this work we introduce a pion exchange and a contact term as
the main contributions to the direct $D^{*0}{D}^{*-}\to
D^{*0}{D}^{*-}$ interaction as listed in the second row of Fig.
\ref{sp} depicted by the white kernel which is included in the grey kernel.
{The contact term can be regarded as an effective one due to the collection of heavier particle exchanges
other than pion and hence there appears a relative phase to one pion exchange term. }
Here, we need to emphasize that we may use only a contact term to construct a much simpler model assuming one pion exchage can be approximated as a contact term. However, one pion exchange actually denotes the long-distant contribution while the contact term reflects the short-distant contrition from the heavier meson exchanges. Considering these facts, we would like to introduce both one pion exchange and contact term.

In the following, we first give the general formula describing the two body $\to$ two body process.

Before obtaining a total amplitude of Fig. \ref{Fig:main} and all the ladder diagrams of the first row of
Fig. \ref{sp}, we need to consider the following items.
\begin{enumerate}
\item Feynman diagrams in Fig. \ref{sp} are directly described in terms of two-particle bases,
    $\left|\boldsymbol{p}_1,\boldsymbol{p}_2,s_1,s_2,\sigma_1,\sigma_2\right\rangle$, in the initial
    and final states. On the other hand, a simple relation between grey and white blobs is obtained if
    they are expressed in terms of a partial wave basis, $|p,j,\sigma,\ell,s\rangle$. Detailed
    propertied of these bases which are used in deriving equations below are give in Appendix.
\item Insert cuts in all the places in a grey blob where two propagators connecting two white blobs
    like the first row in Fig. \ref{sp} appear.
\item Expand the grey blob in terms of the white blobs.
\item Attach the tree vertex, $Y(4260)\pi^+D^{*0}D^{*-}$, to the grey blob in Fig. \ref{sp}.
\item One cut is now inserted between the vertex and the grey blob as in Fig. \ref{Fig:cut} so that a
    total amplitude in Fig. \ref{Fig:main} can be a multiplication of the tree vertex,
    $Y(4260)\pi^+D^{*0}D^{*-}$, and the approximate ladder diagrams.
\end{enumerate}

Using Eq. (\ref{pT}), the grey and white blobs are expressed as
\begin{eqnarray}
  \left\langle
  q,\tilde{j},\tilde{\sigma},\tilde{\ell},\tilde{s}\left|T\right|p,j,\sigma,\ell,s\right\rangle &=&
  \mathcal T^{(j)}(p)^{\tilde{\ell},\tilde{s}}_{\ell,s}\delta^4(p-q)~, \\
  \left\langle
  q,\tilde{j},\tilde{\sigma},\tilde{\ell},\tilde{s}\left|T_0\right|p,j,\sigma,\ell,s\right\rangle &=&
  {\mathcal{T} _0^{(j)}(p)}^{\tilde{\ell},\tilde{s}}_{\ell,s}\delta^4(p-q)~, \label{gw}
\end{eqnarray}
which are expressed in partial wave bases and $T$ and $T_0$ are the corresponding $T$ matrices,
respectively.
Following the items 2 and 3 above, the diagrams listed in the first row of Fig. \ref{sp} can be finally
described by the series
\begin{eqnarray}
\mathcal T^{(j)}(p)
&=&1+i\beta\mathcal  T_0^{(j)}(p)
+(i\beta)^2\mathcal  T_0^{(j)}(p) \mathcal T_0^{(j)}(p)+\dots\nonumber\\
&=&\frac{1}{1-i\beta\mathcal T^{(j)}_0(p)}, \label{series}
\end{eqnarray}
where $\beta=(2\pi)^4/2$, all the indices are suppressed and the amplitude
$\mathcal{T}^{(j)}_0(p)^{\tilde{\ell},\tilde{s}}_{\ell,s}$
which
consist of the one-pion exchange contribution and the contact term. All the quantities included in
Eq.~(\ref{series}) should be tacitly understood to be matrices and be accordingly multiplied with each
other.

Thus, by inserting the completeness condition for two-particle bases given by Eq. (\ref{identity1}) we can
express $\mathcal{T}^{(j)}_0(p)_{\ell,s}^{\tilde\ell,\tilde s}$ as,
\begin{eqnarray}
&&\mathcal{T}^{(j)}_0(p)^{\ell,s}_{\tilde\ell,\tilde s}\delta^4(p-q) \nonumber\\
&&\equiv\langle q,\tilde{j},\tilde{\sigma},\tilde{\ell},\tilde{s}|T_0|p,j,\sigma,\ell,s\rangle\nonumber\\
&&=\sum_{s_i,s'_i,\sigma'_i,\sigma_i}
\int d^3\tilde{p}_1d^3\tilde{p}_2d^3\tilde{p}'_1d^3\tilde{p}'_2\nonumber\\
&&\quad\times\langle
q,\tilde{j},\tilde{\sigma},\tilde{\ell},\tilde{s}|\boldsymbol{p}_1,s_1,\sigma_1;\boldsymbol{p}_2,s_2,\sigma_2\rangle\nonumber\\
&&\quad\times\langle\boldsymbol{p}'_1,s'_1,\sigma'_1;\boldsymbol{p}'_2,s'_2,\sigma'_2|p,j,\sigma,\ell,s\rangle\nonumber\\
&&\quad\times\langle\boldsymbol{p}_1,s_1,\sigma_1;\boldsymbol{p}_2,s_2,\sigma_2|T_0|\boldsymbol{p}'_1,s'_1,\sigma'_1;\boldsymbol{p}'_2,s'_2,\sigma'_2\rangle\label{pac}
\end{eqnarray}
with $d^3\tilde p=d^3p/\left((2\pi)^32E_p\right)$.

Having the above preparation and using the effective Lagrangian approach, we illustrate how to obtain a
matrix element
$\langle\boldsymbol{p}_1,s_1,\sigma_1;\boldsymbol{p}_2,s_2,\sigma_2|T_0|\boldsymbol{p}'_1,s'_1,\sigma'_1;\boldsymbol{p}'_2,s'_2,\sigma'_2\rangle$
for the discussed $D^{*0}{D}^{*-}\to
D^{*0}{D}^{*-}$ interaction. The involved effective Lagrangians are given by
\begin{eqnarray}
\mathcal L_{Y(4260)D^{*} D^{*}\pi}&=&-ig_{YD^*D^*\pi}\epsilon^{\mu\nu\rho\sigma}Y_\mu
D^*_\nu\partial_\rho\pi\bar D^*_\sigma\nonumber\\
&&-ih_{YD^*D^*\pi}\epsilon^{\mu\nu\rho\sigma}\partial_\mu Y_\nu D^*_\rho\pi\bar D^*_\sigma,\\
\mathcal L_{D^*D^*D^*D^*}&=&g_{c}\left(2\bar D^{*\dagger}_\nu D^*_\mu\bar D^{*\dagger\nu}D^{*\mu}-\bar
D^{*\dagger}_\mu D^{*\mu}\bar D^{*\dagger}_\nu D^{*\nu}\right.\nonumber\\
&&\quad\left.-\bar D^{*\dagger}_\mu D^*_\nu\bar D^{*\dagger\nu} D^{*\mu}\right),\\
\mathcal L_{D^*D^*\pi}&=&-g_{D^*D^*\pi}\epsilon^{\mu\nu\rho\sigma}\partial_\mu D_\nu^*\pi\partial_\rho\bar
D^*_\sigma,
\end{eqnarray}
where these Lorentz structures are given in Refs. \cite{Kaymakcalan:1984,Haglin:2000,Colangelo:2002mj}.
These effective Lagrangians are
obtained by assuming the $SU(2)$ invariance among couplings of
$SU(2)$ pseudoscalar and vector multiplets as usual. 
{The coupling constant $g_{D^*D^*\pi}$ can be related to the $D^*\to D\pi$ decay \cite{Ahmed:2001xc} and
hence, $g_{D^*D^*\pi}=8.94$ GeV$^{-1}$ is obtained \cite{Isola:2003fh}.}
However, the coupling constants $g_{YD^*D^*\pi}$ and
$h_{YD^*D^*\pi}$ cannot be constrained since they are related to the
inner structure of $Y(4260)$. In this work, we will discuss the line
shapes of the $D^{*0}D^{(*-)}$ invariant mass spectra when taking
different values of $\xi=h_{YD^*D^*\pi}/g_{YD^*D^*\pi}$.

The amplitude for the interaction $D^{*0}(p_2,\epsilon_2)D^{*-}(p_1,\epsilon_1)\rightarrow
D^{*0}(p_4,\epsilon_4)D^{*-}(p_3,\epsilon_3)$ by exchanging one pion is
\begin{eqnarray}
\mathcal{A}_{\pi-exchange}&=&-g^2_{D^*D^*\pi}\varepsilon^{\mu\nu\rho\sigma}p_{1\mu}
\epsilon_{1\nu}p_{3\rho}\epsilon^*_{3\sigma}\frac{1}{q^2-m_\pi^2}\nonumber\\
&&\times\varepsilon^{\alpha\beta\gamma\delta}
p_{4\alpha}\epsilon^*_{4\beta}p_{2\gamma}\epsilon_{2\delta},\label{am1}
\end{eqnarray}
while the amplitude for the contact term
reads as
\begin{eqnarray}
\mathcal
A_{contact}&=&g^2_{D^*D^*D^*D^*}\left[4(\epsilon_{2}\cdot\epsilon_3^{*})(\epsilon_{1}\cdot\epsilon^{*}_4)-2(\epsilon_{2}\cdot\epsilon^{*}_4)(\epsilon_{1}\cdot\epsilon^{*}_3)\right.\nonumber\\
&&\left.-2(\epsilon_{1}\cdot\epsilon_2)(\epsilon^{*}_{3}\cdot\epsilon^{*}_4)\right].\label{am2}
\end{eqnarray}
In addition, the tree amplitude of the direct $Y(4260)(p,\epsilon_Y)\rightarrow
D^{*0}(p_2,\epsilon_2)D^{*-}(p_1,\epsilon_1)\pi^+(k)$ decay is
\begin{eqnarray}
\mathcal A_{YD^*D^*\pi}&=&-g_{YD^*D^*\pi}\varepsilon^{\mu\nu\rho\sigma}
\epsilon^*_{1\mu}\epsilon^*_{2\nu}\epsilon_{Y\rho}\nonumber\\&&\times(5k+p_1+p_2+3{\xi}p)_\sigma.\label{am3}
\end{eqnarray}
As for the process $D^{*0}{D}^{*-}\to
D^{*0}{D}^{*-}$, a matrix element
$\langle\boldsymbol{p}_1,s_1,\sigma_1;\boldsymbol{p}_2,s_2,\sigma_2|T_0|\boldsymbol{p}'_1,s'_1,\sigma'_1;\boldsymbol{p}'_2,s'_2,\sigma'_2\rangle$
in Eq. (\ref{pac}) can be further expressed as
\begin{eqnarray}
&&\langle\boldsymbol{p}_1,s_1,\sigma_1;\boldsymbol{p}_2,s_2,\sigma_2|T_0|\boldsymbol{p}'_1,s'_1,\sigma'_1;\boldsymbol{p}'_2,s'_2,\sigma'_2\rangle\nonumber\\
&&=\mathcal A_{\pi-exchange}+e^{i\phi}\mathcal A_{contact}, \label{sum_cpi}
\end{eqnarray}
{where the phase $\phi$ is introduced.}
{Following the item 3, we need to insert the cut in between the tree vertex and the grey blob as in Fig.
\ref{Fig:cut}. There have been a couple of examples to calculate the decay amplitudes by inserting a cut
between a tree vertex and other diagrams. See, e.g., Refs. \cite{Chen:2011yu,Chen:2010an} and
\cite{Chen:2008ee}.}
Finally, by using Eq. (\ref{series}), the total partial wave amplitude for the process $Y(4260)\to
\pi^+D^{*0}D^{*-}$ discussed in this work becomes
\begin{eqnarray}
\mathcal T_{\text{total}}^{(j)}(p)_{\ell,s}^{\tilde \ell,\tilde s}&=&\mathcal T^{(j)}(p)^{\tilde
\ell,\tilde s}_{\ell',s'}\mathcal T^{(j)}_{YD^*D^*\pi}(p)^{\ell',s'}_{\ell,s},
\end{eqnarray}
where $\mathcal T^{(j)}_{YD^*D^*\pi}(p)^{\ell',s'}_{\ell,s}$ is the tree amplitude given by Eq.
(\ref{am3}) expressed in partial wave bases like in Eq. (\ref{pac}).

Summing over all $\mathcal T_{\text{total}}^{(j)}(p)_{\ell,s}^{\tilde \ell,\tilde s}$~ partial amplitudes
with different quantum numbers, we get the total amplitude $\mathcal{M}$. The differential decay width
reads as
\begin{eqnarray}
d\Gamma=\frac{1}{(2\pi)^516M_{Y(4260)}^2}|\mathcal
M|^2|\boldsymbol{p}_{D^*}^*||\boldsymbol{p}_\pi|dm_{D^*\bar{D}^*}d\Omega^*_{D^*}d\Omega_\pi,
\end{eqnarray}
where $\boldsymbol{p}_\pi$ is a three-momentum of the emitted pion in the rest frame of $Y(4260)$, while
$(\boldsymbol{p}_{D^*}^*, \Omega_{D^*}^*)$ is a momentum and angle of $D^*$ in the cms rest frame of the
$D^{*0}$ and $D^{*-}$ mesons. $\Omega_\pi$ denotes the angle of $\pi$ in the rest frame of $Y(4260)$ and
$m_{D^*\bar{D}^*}$ is the $D^{*0}D^{*-}$ invariant mass.

\section{Numerical Results}\label{sec3}

\begin{figure}[htb]
\begin{tabular}{c}
\includegraphics[width=0.48\textwidth]{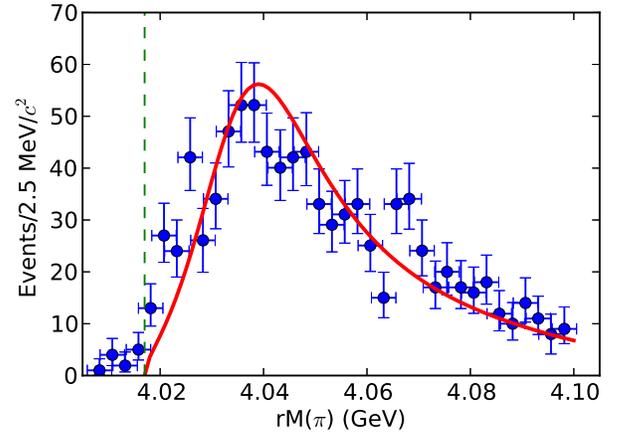}
\end{tabular}
\caption{(Color online.)
{The obtained line shape (red curve) in the $D^*\bar{D}^*$ invariant mass spectrum of $Y(4260)\to\pi^+
D^{*0}{D}^{*-}$ and the comparison with the experimental data (blue dots with error)
\cite{Ablikim:2013nra}. Here, the green vertical dashed line denote the $D^*\bar{D}^*$ threshold. Our
theoretical result (red curve) is obtained by taking the typical values $g_c=60.827$, $\phi=0.986$ and
$\xi=1$.}}\label{Fig:result}
\end{figure}

With the above analytical calculations, in the following we present the differential decay width of
$Y(4260)\to\pi^+ D^{*0}D^{*-}$ dependent on the $m_{D^*\bar{D}^*}$
invariant mass (see Fig. \ref{Fig:result}), where three free parameters $g_c$, $\xi$ and $\phi$ are
involved in our calculation. By this study, we want to answer whether the newly observed $Z_c(4025)$
can be reproduced by our model. {In our numerical calculation, we only take the $\ell=0$ partial wave since its contribution is dominant.}

The comparison between our theoretical result and the experimental data indicates that we can simulate
the enhancement structure near the $D^*\bar{D}^*$ threshold, which is similar to the $Z_c(4025)$ structure
observed by BESIII just shown in Fig. \ref{Fig:result}.  Here, we notice that there appear experimental
data below the threshold, which is due to the adopted experimental method, i.e., BESIII has studied the
$\pi^\mp$ recoil mass spectrum \cite{Ablikim:2013nra}.
The above comparison between theoretical and experimental results provides a direct evidence that  newly
observed $Z_c(4025)$ structure can be well understood via the ISPE mechanism.

\section{Summary}

In summary, a new charged enhancement $Z_c(4025)$ near the $D^*\bar{D}^*$ threshold has been reported by
BESIII. This intriguing experimental observation not only makes the family of the charged charmoniumlike
structure become abundant, but also stimulates our interest in revealing the underlying mechanism behind
this novel phenomenon.
In this work, we study the $Y(4260)\to\pi^+ D^{*0}D^{*-}$ decay via the ISPE mechanism, where the involved
$D^{*0}D^{*-}\to D^{*0}D^{*-}$ interaction is considered by introducing the ladder diagrams.
Our result shows that there exists an enhancement structure near $D^*\bar{D}^*$ threshold appearing in
the $D^*\bar{D}^*$ invariant mass spectrum of $Y(4260)\to\pi^+ D^{*0}D^{*-}$, which can correspond to the
newly observed $Z_c(4025)$. This fact indicates that the ISPE mechanism existing in the $Y(4260)$ decays
can be as one of the possible mechanisms to explain this new BESIII's observation.

At present, experiment has made big progress on searching for charged bottomoniumlike and charoniumlike
structures. Studying these phenomena is an interesting research topic with full of challenges and
opportunities. Further theoretical and experimental efforts will be helpful to finally understand what is
the reason resulting in these observations.

{Before closing this section, we need to discuss further developments of 
our model:}

{1. In this work, we have neglected the coupled-channel effect arising from one-loop box diagrams like $D^{*0} D^{*-}\to D^{0} D^{-}/D^{*0}D^{-}/D^0 D^{*-}\to D^{*0} D^{*-}$ via two $\pi^0$ exchanges. Hence, in the next step, we need to include the coupled-channel effect in our model.}

{2. In this work, we introduce only the imaginary part of the loop when calculating the diagrams listed in Fig. \ref{sp}. To some extent, this treatment is an approximation. Hence, we need to develop our model to include the real part of loop diagrams.}


\vfil

\section*{Acknowledgement}

This project is supported by the National Natural Science
Foundation of China under Grants No. 11222547, No. 11175073, No.
11005129 and No. 11035006, the Ministry of Education of China
(FANEDD under Grant No. 200924, SRFDP under Grant No.
20120211110002 , NCET, the Fundamental Research Funds for the
Central Universities), the Fok Ying Tung Education Foundation (No.
131006), and the West Doctoral Project of the Chinese Academy of
Sciences.

\vfil
\appendix
\section{Relation between partial wave and two-particle bases}

A two-particle state,
which is characterized by the three-momenta $\boldsymbol{p}_1,\boldsymbol{p}_2$ and the $z$-components
$\sigma_1,\sigma_2$ of the corresponding spins $s_1,s_2$, can be defined as
$\left|\boldsymbol{p}_1,\boldsymbol{p}_2,s_1,s_2,\sigma_1,\sigma_2\right\rangle$, which satisfies the
normalization
\begin{eqnarray}
&&\langle\boldsymbol{p}'_1,s'_1,\sigma'_1;\boldsymbol{p}'_2,s'_2,\sigma'_2|\boldsymbol{p}_1,s_1,\sigma_1;\boldsymbol{p}_2,s_2,\sigma_2\rangle\nonumber\\
&&=\tilde{\delta}^{3}(\boldsymbol{p}_1-\boldsymbol{p}'_1)\tilde{\delta}^{3}(\boldsymbol{p}_2-\boldsymbol{p}'_2)
\delta_{s'_1s_1}\delta_{s'_2s_2}\delta_{\sigma_1\sigma'_1}\delta_{\sigma_2\sigma'_2}
\end{eqnarray}
with $\tilde{\delta}^{3}(\boldsymbol{p})=(2\pi)^32E\delta^{3}(\boldsymbol{p})$. The completeness condition
gives us,
\begin{eqnarray}
&&\sum_{s_1,s_2,\sigma_1,\sigma_2 }\int
d^3\tilde{p}_1d^3\tilde{p}_2\left|\boldsymbol{p}_1,s_1,\sigma_1;\boldsymbol{p}_2,s_2,\sigma_2\right\rangle
\nonumber\\
&&\quad\times\left\langle\boldsymbol{p}_1,s_1,\sigma_1;\boldsymbol{p}_2,s_2,\sigma_2\right|=1,\label{identity1}
\end{eqnarray}
where $d^3\tilde{p}=d^3p\big/\big((2\pi)^32E_p\big)$.

A partial wave base can be expressed in terms of two-particle bases as
\begin{eqnarray}
&&|p,j,\sigma,\ell,s\rangle\nonumber\\&&=\frac{1}{2(2\pi)^3}\sqrt{\frac{p}{E_p}}\sum_{m,\ell_z}(s_1,\sigma_1;s_2,\sigma_2|s,m)(s,m;\ell,\ell_z|j,\sigma)\nonumber\\
&&\quad\times\int{ d\Omega_{p_1}}
Y_\ell^{\ell_z}(\Omega_{p_1})|\boldsymbol{p}_1,s_1,\sigma_1;\boldsymbol{p}_2,s_2,\sigma_2\rangle,\label{pwb}
\end{eqnarray}
where $\Omega_{p_1}$ is a solid angle of a momentum $\boldsymbol{p}_1$, $Y_\ell^{\ell_z}(\Omega_{p_1})$
denotes spherical harmonics, $\vec{j}=\vec{\ell}+\vec{s}$ and $\vec{s}=\vec{s}_1+\vec{s}_2$. We also
define four-momentum $p=p_1+p_2$ and $E_p$ is a zeroth component of $p$. In addition,
$(s_1,\sigma_1;s_2,\sigma_2|s,m)$ and $(s,m;\ell,\ell_z|j,\sigma)$ are the Clebsch-Gordan coefficients.
A partial wave base satisfies the normalization
\begin{equation}
  \langle p',j',\sigma',\ell',s'|p,j,\sigma,l,s\rangle =
  \delta^4(p'-p)\delta_{j'j}\delta_{\sigma'\sigma}\delta_{\ell'\ell}\delta_{ss'}.
\end{equation}
The completeness condition gives us,
\begin{equation}
\sum_{j,\sigma, \ell, s}\int d^4p|p,j,\sigma,\ell,s\rangle\langle p,j,\sigma,\ell,s|=1.\label{e1}
\end{equation}
The inner product of two different bases is given by
\begin{eqnarray}
&&\langle\textbf{p}_1,s_1,\sigma_1;\textbf{p}_2,s_2,\sigma_2|p,j,\sigma,\ell,s\rangle\nonumber\\
&&=2(2\pi^3)\sqrt{E\over{|\boldsymbol{p}_1|}}\,\delta^{4}(p_1+p_2-p)\nonumber\\
&&\times\sum_{m,\ell_z}(s_1,\sigma_1;s_2,\sigma_2|s,m)(s,m;\ell,\ell_z|j,\sigma)Y_\ell^{\ell_z}(\Omega_{p_1}).
\end{eqnarray}

Any amplitude can be expressed as a tensor form in terms of partial wave bases
\begin{eqnarray}
\langle q,\tilde{j},\tilde{\sigma},\tilde{\ell},\tilde{s}|T|p,j,\sigma,\ell,s\rangle=\mathcal
T^{(j)}(p)^{\tilde{\ell},\tilde{s}}_{\ell,s}\delta^4(p-q)~.\label{pT}
\end{eqnarray}
Here, we need to notice that Eq. (\ref{pT}) is independent on the quantum numbers $\tilde{j}$,
$\tilde\sigma$ and $\sigma$, which
is consistent with the constraint from the Wigner-Eckart theorem.


\end{document}